\documentclass[preprint,12pt]{elsarticle}

\usepackage{graphicx}

\usepackage{amssymb}

\usepackage{lineno}

\begin{document}
\begin{frontmatter}



\title{Covariant Formulation of the Transition Radiation Energy Spectrum of an Electron Beam at a Normal Angle of Incidence onto a Round Metallic Screen}

\author{Gian Luca Orlandi}
\ead{gianluca.orlandi@psi.ch}
\address{Paul Scherrer Institut, 5232 Villigen PSI, CH.}

\begin{abstract}
In the transition radiation emission from a $N$ electron bunch
hitting at a normal angle of incidence a metallic screen, the
transverse and the longitudinal spatial coordinates of the electron
bunch play different roles in determining the $N$ single electron
radiation field amplitudes and their relative phases in relation to
the different physical constraints which an electromagnetic
radiative mechanism by a charged beam must meet: i.e., temporal
causality and covariance. The distribution of the $N$ electron
longitudinal coordinates determines indeed the sequence of the $N$
electron collisions onto the metallic screen and, on the basis of
the temporal causality principle, it also determines the
distribution function of the relative emission phases of the $N$
single electron field amplitudes from the metallic surface. The
distribution of the transverse coordinates of the $N$ electrons
contributes as well to determine the relative phase distribution of
the $N$ electron field amplitudes at the observation point - located
on the longitudinal axis of the reference frame - providing a
further phase information that accounts for the transverse
displacement of the $N$ electrons with respect to the beam axis. The
distribution of the transverse coordinates of the $N$ electrons is a
relativistic invariant under a Lorentz transformation with respect
to the direction of motion of the beam and, consequently, it is
expected to leave a covariant mark on the $N$ single electron
amplitudes composing the radiation field. The covariant imprinting
of the $N$ electron transverse density on the radiation field
affects both the temporal coherent and incoherent parts of the
transition radiation energy spectrum. Such a dependence of the $N$
single electron radiation field amplitudes on the electron density
in the transverse plane manifests itself as an increase - towards an
asymptotic limit - of the radiated spectral energy with the decrease
of the beam transverse size and as a spectral narrowing of the
angular distribution of the radiation intensity with respect to the
ideal case of a single electron hitting an infinite metallic screen.
In the case of a round metallic screen with an arbitrary radius, the
formal expression of the transition radiation energy spectrum will
be derived and numerical results will be presented.
\end{abstract}

\begin{keyword}
Form Factor \sep Coherence \sep Fourier Transform \sep Collective
Effects
\PACS 41.60.-m \sep 41.75.-i \sep  42.25.Kb \sep  42.30.Kq
\end{keyword}
\end{frontmatter}

\linenumbers

\section{Introduction}

A relativistic charge crossing a dielectric interface in a
rectilinear and uniform motion can originate a highly directional
and broad wavelength band radiation emission propagating backward
and forward from the boundary surface within a small angle scaling
down with the energy of the charge, the so called transition
radiation \cite{gifr,gari,gari2,frank2,bass,frank,ter,ginz,jack}.
Thanks to the instantaneous, highly directional and charge-energy
dependent features, transition radiation is a precious tool in the
beam diagnostics of a particle accelerator. In the vacuum chamber of
a particle accelerator the relativistic charged beam can be indeed
intercepted by a movable thin metallic screen and the resulting
light pulse can be imaged by a CCD-camera (Charge-Coupled-Device),
for instance, to measure the transverse profile of the charged beam.
Under the key-word OTR Beam Diagnostics (Optical Transition
Radiation), the reader can find an enormous specialized literature
dedicated to this topic. For most part of the practical detection
conditions (from the extreme visible to the THz region), the
metallic screen can be assimilated to an ideal conductor surface and
the transition radiation emission can be easily schematized as the
result of the dipolar oscillation of the conduction electrons of the
metallic surface induced by the electromagnetic field of the
incident relativistic charge. The dipolar oscillation induced in the
double layer of charge by the relativistic charge explains indeed
how an electromagnetic radiative mechanism, originated by a charge
in a rectilinear and uniform motion, can also propagate in the
backward direction. Such a model of the double layer of charge,
describing the transition radiation emission as the result of the
dipolar oscillation of the conduction electrons induced by the
relativistic charge, is also enlightening about the kinematics of
this radiative mechanism and about the common relativistic nature
that it shares with other electromagnetic radiative mechanisms by
relativistic charged beams. The kinematics of the transition
radiative mechanism can be indeed schematized as the head-on
collision of two distributions of charge as observed in the
reference frame of rest of one of the two colliding charged
distributions. The backward and forward double conical transition
radiation emission can be thus interpreted as the photon
bremsstrahlung emission that two head-on colliding electron beams
can originate. Taking into consideration that the transition
radiation mechanism shares with other electromagnetic radiative
mechanisms, such as the synchrotron or the bremsstrahlung radiation,
the same kinematical and relativistic nature, it is thus reasonable
to expect that, even at a very short wavelength, some spectral
modifications of the radiation intensity due to transverse density
of the beam should also affect the transition radiative mechanism by
an electron beam in a similar way as, in other electromagnetic
radiative mechanism by charged beams, the beam transverse size
contributes to determine the so called Brilliance or Luminosity
properties of the radiation source.

In the case of an electron beam at a normal angle of incidence onto
a metallic screen with arbitrary size and shape, it can be
demonstrated that a covariant and temporal causal formulation of the
transition radiation energy spectrum of an electron beam necessarily
implies, even at a very short wavelength, a dependence of the
radiation spectral intensity on the distribution function of the
particle density in the transverse plane \cite{giancov}. In fact, in
the transition radiation emission of an electron beam at a normal
angle of incidence onto a metallic screen, the longitudinal and
transverse coordinates of the electrons play different roles. The
distribution function of the longitudinal coordinates of the $N$
electrons determines indeed the sequence of the particle collisions
onto the metallic screen and, consequently, on the basis of the
temporal causality principle, also the distribution function of the
relative emission phases - from the metallic surface - of the $N$
single electron amplitudes composing the radiation field. The
distribution function of the $N$ electron transverse coordinates
contributes as well in determining the relative phase delay of the
$N$ single electron field amplitudes at the observation point as a
function of the transverse displacement of the $N$ electrons with
respect to the beam axis where the radiation field is supposed to be
observed. But, in addition, because of the relativistic invariance
under a Lorentz transformation in the direction of motion of the
beam, the $N$ electron transverse density manifests itself as a
covariant feature of the radiation field whose observability can
transform but not disappear under a Lorentz transformation. In the
present work, the covariant and temporal causal formula of the
transition radiation energy spectrum of a $N$ electron bunch,
already derived in an implicit form in \cite{giancov} in the most
general case of a radiator surface with an arbitrary size and shape,
will be here rendered into an explicit form in the particular case
of a round radiator with an arbitrary radius. The so obtained
formula of the transition radiation energy spectrum of a $N$
electron bunch at a normal angle of incidence onto a round metallic
screen is demonstrated to be temporal causal and covariant
\cite{giancov} and to reproduce, as a limit, some results already
well known in literature, such as: the Frank-Ginzburg formula of the
single particle transition radiation energy spectrum in the ideal
case of an infinite radiator or the single electron transition
energy spectrum radiated by a round metallic screen with a finite
radius. Finally, because of the covariant dependence of the
radiation field on the transverse density of the $N$ electrons, it
results that the distribution function of the $N$ electron
transverse coordinates affects both the temporal coherent and
incoherent parts of the transition radiation energy spectrum. Even
at a very short wavelength, the effect of the $N$ electron
transverse density on the spectral distribution of the radiation
intensity manifests itself as an increase of the radiated energy
with the decrease of the beam transverse size and a spectral
narrowing of the angular distribution of the radiation intensity in
comparison with the ideal case of an electron beam having a
point-like transverse extension. In the following, in the case of a
bunch of $N$ electrons hitting the metallic screen at a normal angle
of incidence, details on the formula of the transition radiation
energy spectrum and numerical results will be presented.

\section{Transition radiation energy spectrum of a $N$ electron bunch}

A bunch of $N$ electrons in a rectilinear and uniform motion along
the $z$-axis of the laboratory reference frame is colliding, at a
normal angle of incidence, onto a flat ideal conductor surface $S$
placed in the plane $z=0$. The $N$ electrons are supposed to fly in
vacuum with a common velocity $\vec{w}=(0,0,w)$. All the electrons
are supposed to hit the metallic screen at a normal angle of
incidence. Effects of the angular divergence of the electron beam on
the radiation energy spectrum are not considered in the present
work. The radiator surface $S$ has a round shape with a finite
radius $R$. The observation point of the radiation emission is on
the $z$-axis at a distance $D$ from the screen much larger than the
observed wavelength $\lambda$. The spatial coordinates of the $N$
electrons being $[\vec{\rho}_{0j}=(x_{0j},y_{0j}),z_{0j}]$ with
$j=1,..,N$ at a given reference time $t=0$ when the center of mass
of the charged distribution is crossing the boundary surface, the
spectral component of the radiation field resulting from the
collision of the $N$ relativistic electrons onto the metallic screen
reads, see
\cite{giancov,gian,gianbis,gianter,gianquater,gian5,gian6},
\begin{eqnarray}
E_{x,y}^{tr}(\vec{\kappa},\omega)=
\sum_{j=1}^{N}H_{x,y}(\vec{\kappa},\omega,\vec{\rho}_{0j})\,e^{-i(\omega/w)z_{0j}}\label{uno}
\end{eqnarray}
where, under the far-field approximation \cite{ter,jack,born}, the
single electron contribution to the radiation field can be
calculated in the most general case of a radiator surface $S$ with
an arbitrary shape and size (either infinite $S=\infty$ or finite
$S<\infty$) as follows
\cite{giancov,gian,gianbis,gianter,gianquater,gian5,gian6}:
\begin{eqnarray}
H_{x,y}(\vec{\kappa},\omega,\vec{\rho}_{0j})=\frac{iek}{2\pi^2Dw}\int\limits_S
d\vec{\rho}\int d\vec{\tau}
\frac{\tau_{x,y}\,e^{-i\vec{\tau}\cdot\vec{\rho}_{0j}}}{\tau^2+\alpha^2}
e^{i(\vec{\tau}-\vec{\kappa})\cdot\vec{\rho}}.\label{due}
\end{eqnarray}
In previous Equation, $k=\omega/c=2\pi/\lambda$ is the wave
number, $\vec{\kappa}=(k_x,k_y)=k\sin\theta(\cos\phi,\sin\phi)$ is
the transverse component of the wave-vector,
$\alpha=\frac{\omega}{w\gamma}$ ($\gamma$ being the relativistic
Lorentz factor) and the vector $\vec{\rho}=(x,y)$ represents the
integration variable on the surface $S$ of the screen.

With reference to Eqs.(\ref{uno},\ref{due}), the transition
radiation energy spectrum by a $N$ electron beam can be finally
calculated as the flux of the Poynting vector, see also
\cite{giancov}:
\begin{eqnarray}
\frac{d^2I}{d\Omega
d\omega}=\frac{cD^2}{4\pi^2}\sum_{\mu=x,y}\left(\sum_{j=1}^{N}\left|H_{\mu,j}
\right|^2+\sum_{j,l(j\neq l)=1}^{N}e^{-i(\omega/w)(z_{0j}-z_{0l})}
H_{\mu,j}H^*_{\mu,l}\right),\label{tre}
\end{eqnarray}
where $H_{\mu,j}=H_{x,y}(\vec{\kappa},\omega,\vec{\rho}_{0j})$ with
$\mu=x,y$, as defined in Eq.(\ref{due}).

The formal expression of the radiation field and of the radiation
energy spectrum of the $N$ electron bunch, as represented in
Eqs.(\ref{uno},\ref{due}) and Eq.(\ref{tre}) in the most general
case of a flat radiator surface $S$ with an arbitrary size and
shape, meets the constraints of the temporal causality and of the
covariance. The structure of the emission phases from the screen of
the $N$ single electron field amplitudes composing the radiation
field - see Eqs.(\ref{uno},\ref{due}) - is indeed causally related
to the temporal sequence of the $N$ particle collision onto the
metallic screen. Furthermore, with reference to \cite{giancov},
covariant or covariance consistent are all the formal steps which,
from the expression of the electromagnetic field of the $N$ electron
bunch in the laboratory reference frame, leads to the radiation
field - see Eqs.(\ref{uno},\ref{due}) - and to the radiation energy
spectrum, see Eq(\ref{tre}). For further details about the covariant
and the temporal causal formulation of the transition radiation
emission from a $N$ electron bunch - as described in
Eqs.(\ref{uno},\ref{due},\ref{tre}) - the reader is addressed to
\cite{giancov}.

For the sake of ease, in order to proceed with the formal explicit
derivation of the transition radiation energy spectrum in the case
of a round radiator with a radius $R$, a polar angle representation
is considered for the following vectors
\begin{displaymath} \left\{
\begin{array}{l}
\vec{\tau}=\tau(\cos\eta,\sin\eta)\\
\vec{\xi}_{j}=(\vec{\rho}-\vec{\rho}_{0j})=\xi_{j}(\cos\psi_{j},\sin\psi_{j})\\
 \end{array} \right.\label{tre-bis}
\end{displaymath}
where
$\xi_{j}=\sqrt{\rho^2+\rho^2_{0j}-2\vec{\rho}\cdot\vec{\rho}_{0j}}\,$
$(j=1,..,N)$, $\vec{\rho}_{0j}=(x_{0j},y_{0j})$ and
$\vec{\rho}=(x,y)$ with $0<\rho<R$ being the vectors of the $N$
electron transverse coordinates and the vector of the screen
coordinates, respectively. Taking into account the following
integral representation of the Bessel function of the first kind
\cite{gradryz} (see also Eq.($17$) in \cite{verzi}):
\begin{eqnarray}
\int\limits_{0}^{2\pi}d\psi \left( \begin{array}{c}
\cos\psi\\
\sin\psi\\
\end{array}\right)\,e^{iab\cos(\psi-\phi)}=2\pi i\label{quattro}\left( \begin{array}{c}
\cos\phi\\
\sin\phi\\
\end{array}\right)J_1(ab)
\end{eqnarray}
in the limit of a round metallic surface with a radius $R$ much
longer than the average transverse size of the beam
$(R\gg<\rho_{0j}>, j=1,..,N)$ the integral that, in
Eq.(\ref{due}), determine the single electron field amplitude can
be explicitly calculated:
\begin{eqnarray}
&&\int\limits_S d\vec{\rho}\int d\vec{\tau}
\frac{\tau_{x,y}\,e^{-i\vec{\tau}\cdot\vec{\rho}_{0j}}}{\tau^2+\alpha^2}
e^{i(\vec{\tau}-\vec{\kappa})\cdot\vec{\rho}}=\nonumber\\
&&=\int\limits_S
d(\vec{\rho}-\vec{\rho}_{0j})\,e^{-i\vec{\kappa}\cdot(\vec{\rho}-\vec{\rho}_{0j}+\vec{\rho}_{0j})}
\int d\vec{\tau}
\frac{\tau_{x,y}\,e^{i\vec{\tau}\cdot(\vec{\rho}-\vec{\rho}_{0j})}}{\tau^2+\alpha^2}=\label{cinque}\nonumber\\
&&=\,e^{-i\vec{\kappa}\cdot\vec{\rho}_{0j}}\int
d\vec{\xi}_{j}\,e^{-i\vec{\kappa}\cdot\vec{\xi}_{j}}\int d\vec{\tau}
\frac{\tau_{x,y}\,
e^{i\vec{\tau}\cdot\vec{\xi}_{j}}}{\tau^2+\alpha^2}=\nonumber\\
&&=\,e^{-i\vec{\kappa}\cdot\vec{\rho}_{0j}}\int
d\vec{\xi}_{j}\,e^{-i\vec{\kappa}\cdot\vec{\xi}_{j}} \int
\frac{d\eta\,d\tau\,\tau^2}{\tau^2+\alpha^2}\left(
\begin{array}{c}
\cos\eta\\
\sin\eta\\
\end{array}\right)\,e^{i\tau\xi_j\cos(\eta-\psi_j)}=\nonumber\\
&&=(2\pi i)e^{-i\vec{\kappa}\cdot\vec{\rho}_{0j}}\int
d\xi_j\,\xi_j\,d\psi_j\left(\begin{array}{c}
\cos\psi_j\\
\sin\psi_j\\
\end{array}\right)e^{-i\kappa\xi_j\cos(\psi_j-\phi)}\int\limits_{0}^{\infty}
\frac{d\tau\,\tau^2\,J_1(\tau\xi_j)}{\tau^2+\alpha^2}=\nonumber\\
&&=(2\pi)^2\alpha\left(\begin{array}{c}
\cos\phi\\
\sin\phi\\
\end{array}\right) e^{-i\vec{\kappa}\cdot\vec{\rho}_{0j}}\int\limits_{\rho_{0j}}^{R+\rho_{0j}}
d\xi_j\,\xi_j\,K_1(\alpha\xi_j)J_1(\kappa\xi_j)=\\
&&=(2\pi)^2\left(\begin{array}{c}
\cos\phi\\
\sin\phi\\
\end{array}\right)\frac{\kappa\,e^{-i\vec{\kappa}\cdot\vec{\rho}_{0j}}}{\kappa^2+\alpha^2}
[\rho_{0j}\Phi(\kappa,\alpha,\rho_{0j})-(R+\rho_{0j})\Phi(\kappa,\alpha,R+\rho_{0j})]\nonumber
\end{eqnarray}
with
\begin{eqnarray}
\Phi(\kappa,\alpha,\rho_{0j})=\alpha
J_0(\kappa\rho_{0j})K_1(\alpha\rho_{0j})+\frac{\alpha^2}{\kappa}
J_1(\kappa\rho_{0j})K_0(\alpha\rho_{0j})\label{sei}
\end{eqnarray}
and
\begin{eqnarray}
\Phi(\kappa,\alpha,R+\rho_{0j})=\alpha
J_0[\kappa(R+\rho_{0j})]K_1[\alpha(R+\rho_{0j})]+\frac{\alpha^2}{\kappa}
J_1[\kappa(R+\rho_{0j})]K_0[\alpha(R+\rho_{0j})]\label{sette}
\end{eqnarray}
where Eqs.(\ref{cinque},\ref{sei},\ref{sette}) follows from
Eq.(\ref{quattro}) and from formulas $6.566(2)$, $6.521(4)$,
$8.473(1)$ and $8.486(17)$ in \cite{gradryz}.

In conclusion, with reference to
Eqs.(\ref{cinque},\ref{sei},\ref{sette}), the harmonic component
of the radiation field produced by a $N$ electron bunch hitting a
round radiator surface with a finite radius $R$ reads, see
Eqs.(\ref{uno},\ref{due}),
\begin{eqnarray}
E_{x,y}^{tr}(\vec{\kappa},\omega)&&=\sum_{j=1}^{N}H_{x,y}(\vec{\kappa},\omega,\vec{\rho}_{0j})\,e^{-i(\omega/w)z_{0j}}=
\sum_{j=1}^{N}\frac{2iek}{Dw}\frac{\kappa}{\kappa^2+\alpha^2}
\,e^{-i[(\omega/w)z_{0j}+\vec{\kappa}\cdot\vec{\rho}_{0j}]}\times\nonumber\\\label{settebis}
&&\times\left(\begin{array}{c}
\cos\phi\\
\sin\phi\\
\end{array}\right)[\rho_{0j}\Phi(\kappa,\alpha,\rho_{0j})-(R+\rho_{0j})\Phi(\kappa,\alpha,R+\rho_{0j})],
\end{eqnarray}
where
$\vec{\kappa}=(\kappa_x,\kappa_y)=k\sin\theta(\cos\phi,\sin\phi)$ is
the transverse component of the wave-vector ($k=2\pi/\lambda$). In
some special and relevant cases, the formula of the transition
radiation field given in Eqs.(\ref{sei},\ref{sette},\ref{settebis})
reproduces as a limit some results already well known in literature.

Case$1$. Single electron $(j=1)$ and ($\rho_{0j}=0$, $R=\infty$):
i.e., a single electron moving on the $z$-axis ($x=y=0$), where the
observation point is also located, in collision onto a radiator
having an infinite surface. In the limit $R\rightarrow\infty$ and
$\rho_{0j}\rightarrow 0$, the functions defined in
Eqs.(\ref{sei},\ref{sette}) tend to the following limit values:
\begin{displaymath}
\left\{
\begin{array}{l}
(R+\rho_{0j})\Phi(\kappa,\alpha,R+\rho_{0j})\rightarrow0\\
\rho_{0j}\Phi(\kappa,\alpha,\rho_{0j})\rightarrow 1
 \end{array} \right.
\end{displaymath}
Taking into account the limits above, the following expression of
the single electron radiation field, see
Eqs.(\ref{sei},\ref{sette},\ref{settebis}), follows
\begin{eqnarray}
E_{x,y}^{tr}(\vec{\kappa},\omega)=\frac{2iek}{Dw}\frac{\kappa}{\kappa^2+\alpha^2}\left(\begin{array}{c}
\cos\phi\\
\sin\phi\\
\end{array}\right)\label{otto}
\end{eqnarray}
where $\kappa=k\sin\theta$. Finally, from the Equation above and
Eq.(\ref{tre}), the well known result of the Frank-Ginzburg formula
of the transition radiation energy spectrum in the ideal case of a
single electron hitting an infinite ideal conductor surface can be
obtained \cite{gifr,gari,gari2,frank2,bass,frank,ter,ginz}:
\begin{eqnarray}
\frac{d^2I_{e}}{d\Omega
d\omega}=\frac{(e\beta)^2}{\pi^2c}\frac{sin^2\theta}
{(1-\beta^2cos^2\theta)^2}.\label{nove}
\end{eqnarray}

Case $2$. Single electron $(j=1)$ and ($\rho_{0j}=0$, $R<\infty$):
similar experimental situation as in previous case but finite radius
of the radiator. With reference to the above reported limit of the
expression in Eq.(\ref{sei}) for $\rho_{0j}\rightarrow 0$, the
radiation field reads, see
Eqs.(\ref{uno},\ref{due},\ref{cinque},\ref{sei},\ref{sette}),
\begin{eqnarray}
E_{x,y}^{tr}(\vec{\kappa},\omega)=\frac{2iek}{Dw}\frac{\kappa}{\kappa^2+\alpha^2}\left(\begin{array}{c}
\cos\phi\\
\sin\phi\\
\end{array}\right)\label{dieci}
[1-\alpha R J_0(\kappa R)K_1(\alpha R)-\frac{\alpha^2R}{\kappa}
J_1(\kappa R)K_0(\alpha R)]
\end{eqnarray}
which, in the far-field approximation, represents the radiation
field of a single electron hitting a finite round metallic screen.
In the case of a single electron hitting a round metallic screen
with a finite radius, as a function of the transverse extension
($\gamma\lambda/2\pi$) of the single electron virtual quanta field
compared to the finite radius $R$ of the radiator, the transition
radiation energy spectrum experiences a low frequency diffractive
cut-off. This problem was already tackled in several works
\cite{shulga1,poty,castellano,casalbuoni1,sutter,casalbuoni2}.
Compare Eq.(\ref{dieci}) in the present work with Eqs.($8,9$) in
\cite{casalbuoni2}, for instance.

Finally, from Eqs.(\ref{tre},\ref{settebis}), the explicit
expression of the transition radiation energy spectrum of a $N$
electron bunch colliding onto a round ideal conductor surface with a
finite radius $R$ can be obtained
\begin{eqnarray}
\frac{d^2I}{d\Omega d\omega}=\frac{d^2I_{e}}{d\Omega
d\omega}\left(\sum_{j=1}^{N}|A_j|^2+\sum_{j,l (j\neq
l)=1}^{N}A_jA^*_l\,e^{-i[(\omega/w)(z_{0j}-z_{0l})+\vec{\kappa}\cdot(\vec{\rho}_{0j}-\vec{\rho}_{0l})]}\right)\label{setteter}
\end{eqnarray}
where $\frac{d^2I_{e}}{d\Omega d\omega}$ is the single electron
radiation energy spectrum already defined in Eq.(\ref{nove}) and
\begin{eqnarray}
A_j=\rho_{0j}\Phi(\kappa,\alpha,\rho_{0j})-(R+\rho_{0j})\Phi(\kappa,\alpha,R+\rho_{0j})
\simeq\rho_{0j}\Phi(\kappa,\alpha,\rho_{0j}),\label{settequater}
\end{eqnarray}
the approximation in Eq.(\ref{settequater}) being valid for most
of the experimental situations where diffractive modifications of
the spectral distribution of the radiated energy due to the finite
size of the screen can be neglected.

In the formula of the transition radiation energy spectrum of a $N$
electron beam, as explicitly derived in
Eqs.(\ref{setteter},\ref{settequater}) in the case of a normal
collision onto a round metallic screen with a finite radius $R$, the
double role played by the transverse coordinates of the $N$
electrons appears evident. On the one hand, as a function of the
displacement of the electrons with respect to the beam axis where
the radiation field is supposed to be observed, they contribute to
determine the relative phase delay of the $N$ electron field
amplitudes at the observation point as the presence of the well
known three-dimensional phase factor in the temporal coherent part
of the formula given in Eq.(\ref{setteter}) clearly indicates. On
the other hand, because of the invariance of the transverse density
of the $N$ electrons under a Lorentz transformation in the direction
of motion of the beam, the $N$ single electron radiation field
amplitudes show a covariant dependence on the transverse density of
the $N$ electron beam. As a result, see
Eqs.(\ref{setteter},\ref{settequater}), both the temporal coherent
and the temporal incoherent components of the transition radiation
energy spectrum of a $N$ electron beam differ from the ideal
scenario of a single electron hitting an infinite metallic screen.
It results indeed that, even at a very short wavelength, the
radiation spectral intensity increases with the decrease of the beam
transverse size towards an asymptotic limit as well as the spectral
angular distribution of the radiation experiences a broadening
towards the asymptotic limit given by the well known ideal case of a
single electron hitting an infinite metallic surface.

\begin{figure*}[tb]
   \centering
   \includegraphics*[width=165mm]{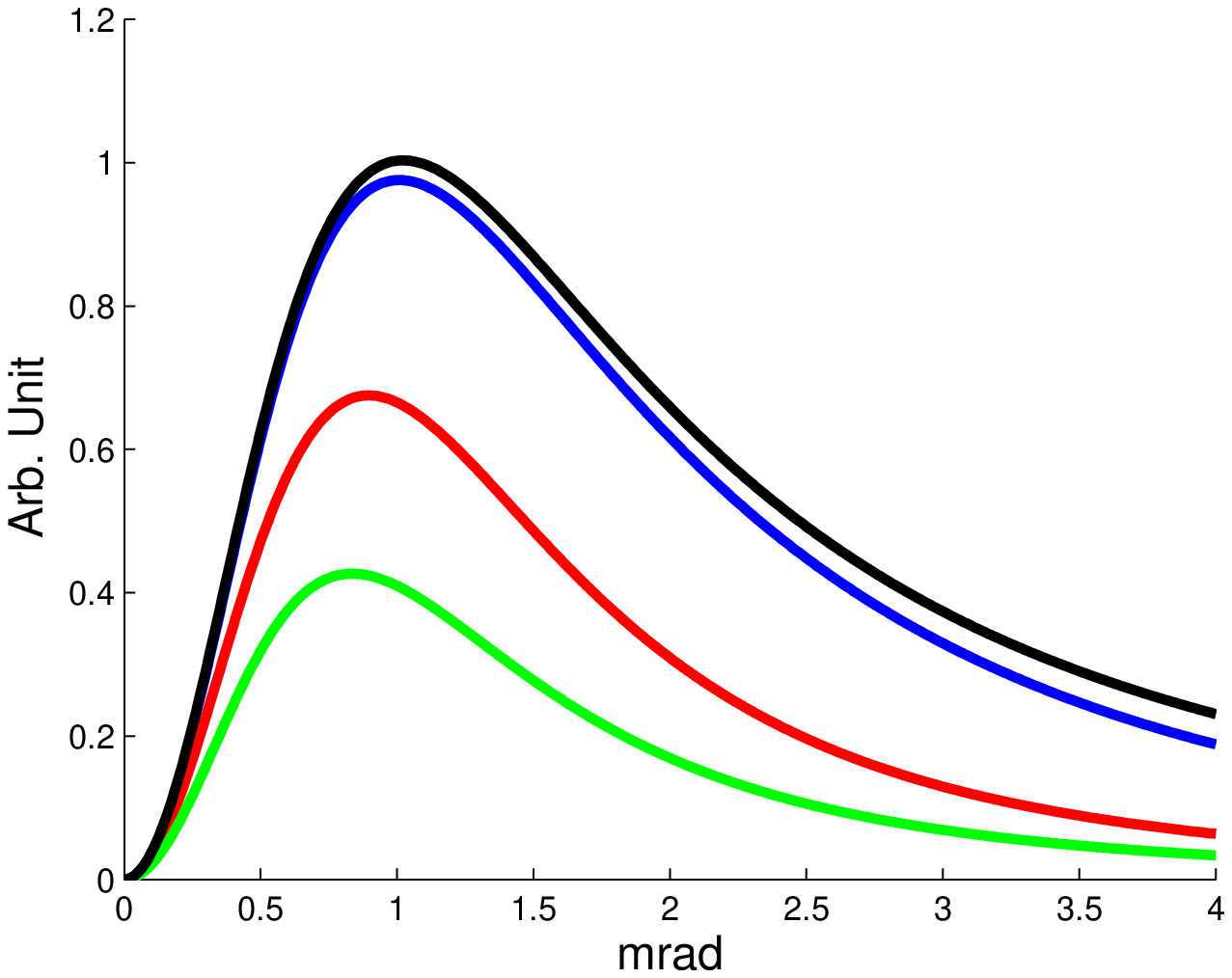}
   \vspace*{-\baselineskip}
   \caption{Angular distribution of the transition radiation emitted at a wavelength $\lambda=530$ $nm$ by a gaussian bunch of $N=10^5$
   electrons and energy $500$ $MeV$ with standard deviation $\sigma=10$ $\mu m$ (Blue curve), $\sigma=50$ $\mu m$ (Red
   curve) and $\sigma=100$ $\mu m$ (Green curve). The Blue, Red and Green curves are
   calculated via Eqs.(\ref{setteter},\ref{settequater}) and compared
   with the analogous result (Black curve) that can be calculated
   in the same conditions of beam energy and
   charge via Eq.(\ref{nove}).}
   \label{Fig1}
\end{figure*}

\begin{figure*}[tb]
   \centering
   \includegraphics*[width=165mm]{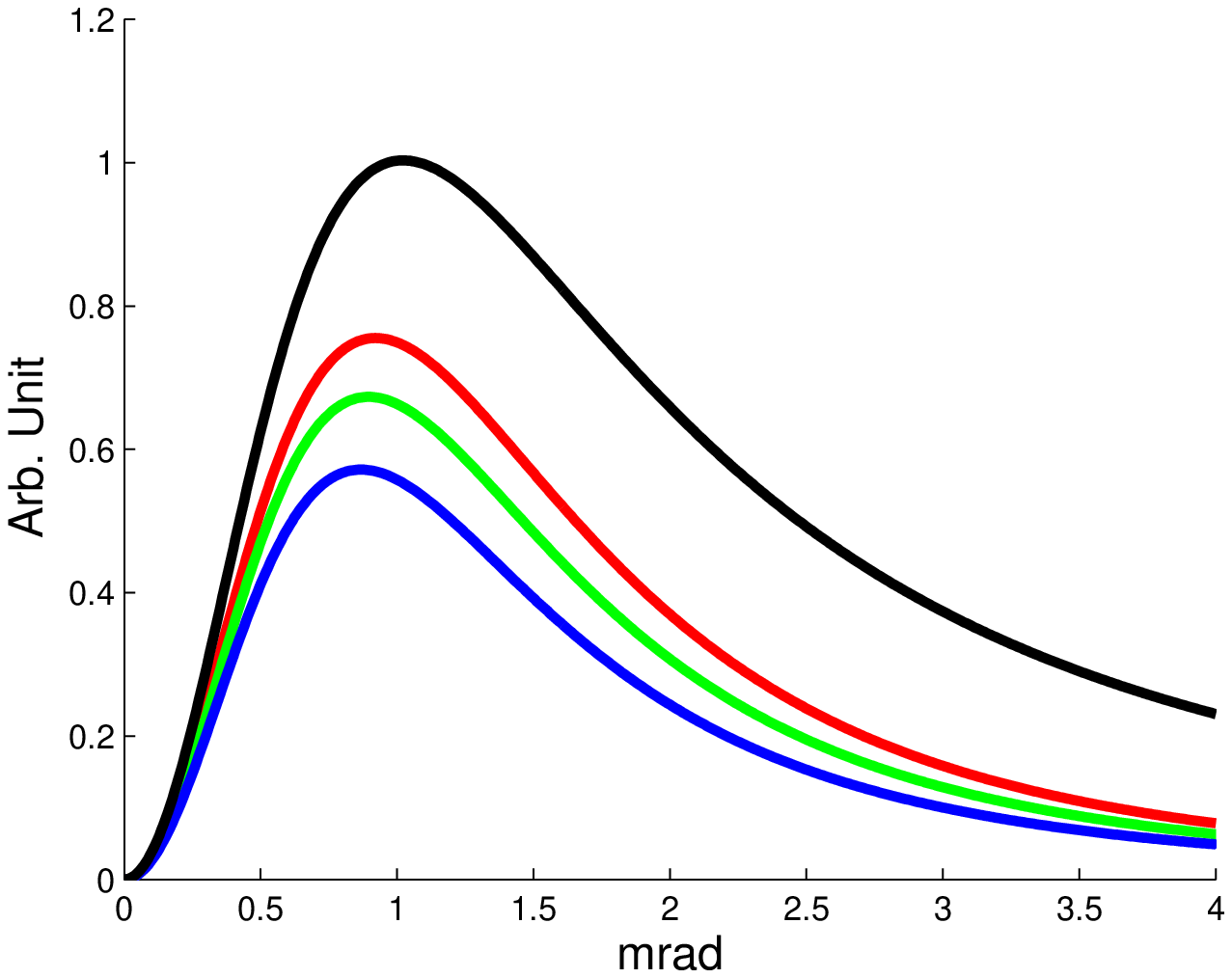}
   \vspace*{-\baselineskip}
   \caption{Beam energy $500$ $MeV$. Angular distribution of the transition radiation emitted at a wavelength $\lambda=680$ $nm$ (Red
   curve), $\lambda=530$ $nm$ (Green curve), $\lambda=400$ $nm$ (Blue curve) by a gaussian bunch of $N=10^5$
   electrons with standard deviation $\sigma=50$ $\mu m$. The Blue, Red and Green curves are
   calculated via Eqs.(\ref{setteter},\ref{settequater}) and compared
   with the analogous result (Black curve) that can be calculated
   in the same conditions of beam energy and
   charge via Eq.(\ref{nove}).}
   \label{Fig2}
\end{figure*}

\begin{figure*}[tb]
   \centering
   \includegraphics*[width=165mm]{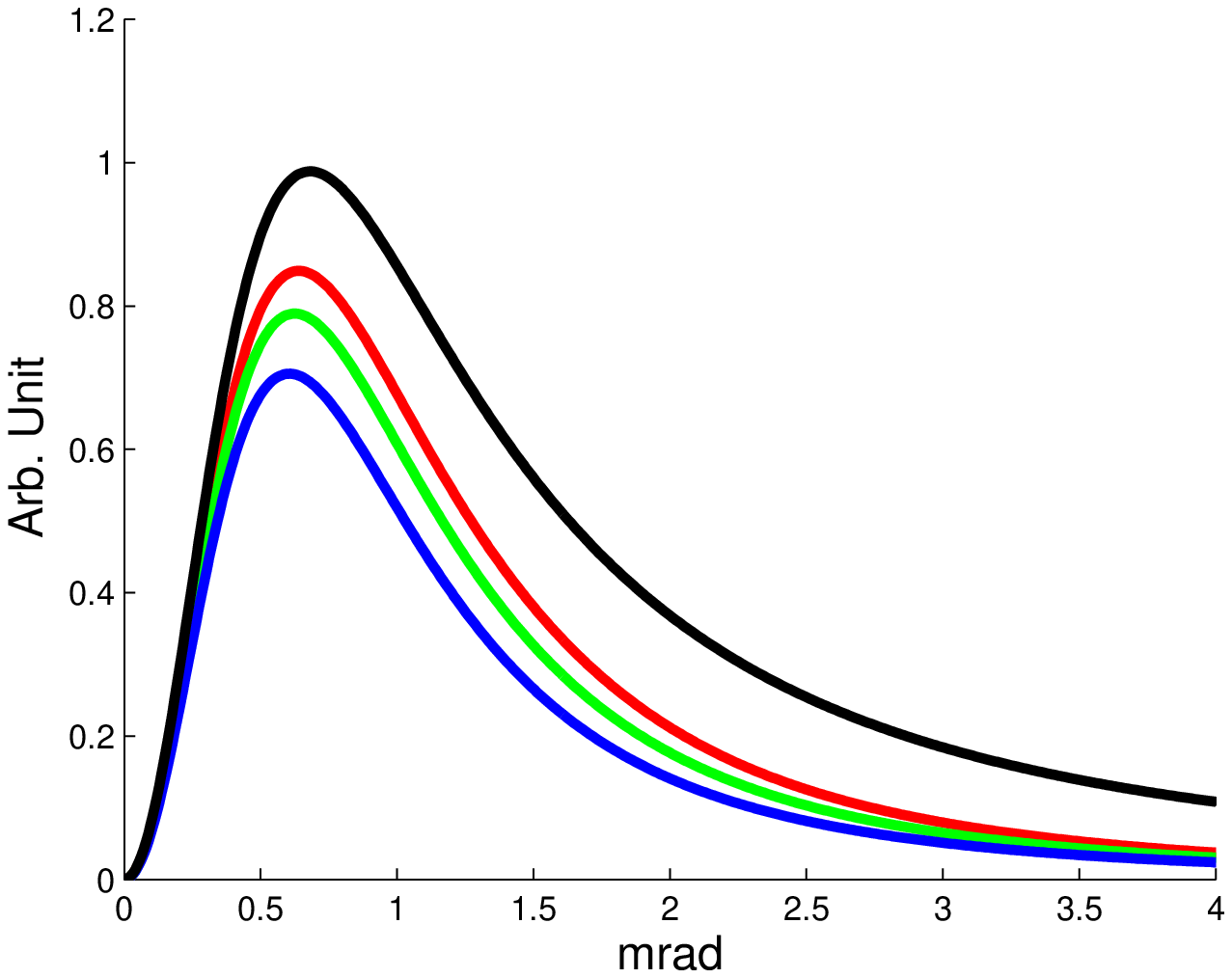}
   \vspace*{-\baselineskip}
   \caption{Beam energy $750$ $MeV$. Angular distribution of the transition radiation emitted at a wavelength $\lambda=680$ $nm$ (Red
   curve), $\lambda=530$ $nm$ (Green curve), $\lambda=400$ $nm$ (Blue curve) by a gaussian bunch of $N=10^5$
   electrons with standard deviation $\sigma=50$ $\mu m$. The Blue, Red and Green curves are
   calculated via Eqs.(\ref{setteter},\ref{settequater}) and compared
   with the analogous result (Black curve) that can be calculated
   in the same conditions of beam energy and
   charge via Eq.(\ref{nove}).}
   \label{Fig3}
\end{figure*}

\begin{figure*}[tb]
   \centering
   \includegraphics*[width=165mm]{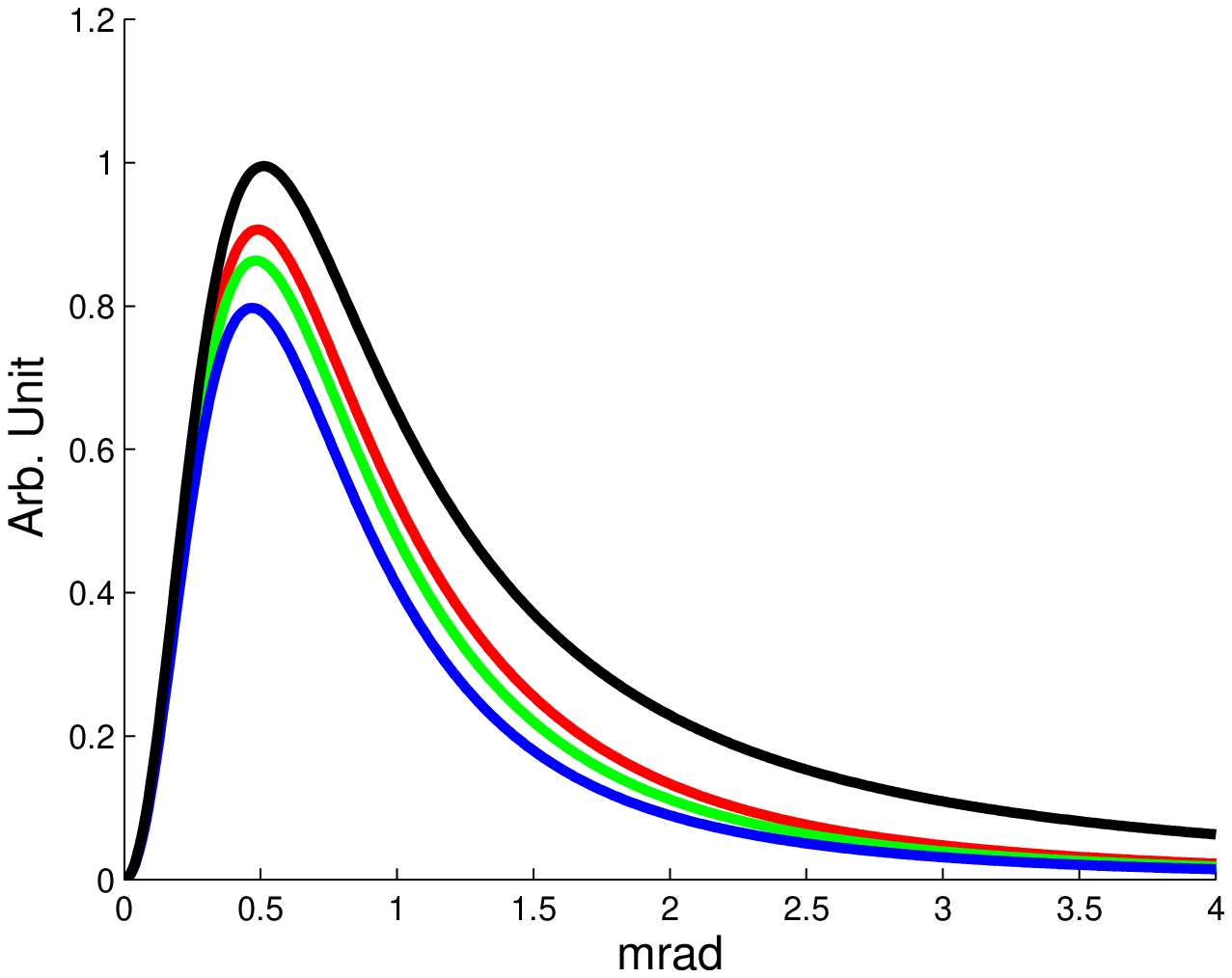}
   \vspace*{-\baselineskip}
   \caption{Beam energy $1000$ $MeV$. Angular distribution of the transition radiation emitted at a wavelength $\lambda=680$ $nm$ (Red
   curve), $\lambda=530$ $nm$ (Green curve), $\lambda=400$ $nm$ (Blue curve) by a gaussian bunch of $N=10^5$
   electrons with standard deviation $\sigma=50$ $\mu m$. The Blue, Red and Green curves are
   calculated via Eqs.(\ref{setteter},\ref{settequater}) and compared
   with the analogous result (Black curve) that can be calculated
   in the same conditions of beam energy and
   charge via Eq.(\ref{nove}).}
   \label{Fig4}
\end{figure*}

In Figs.(\ref{Fig1},\ref{Fig2},\ref{Fig3},\ref{Fig4}), results of
the numerical simulation of the temporal incoherent part of the
transition radiation spectrum produced by a gaussian electron bunch
with a transverse size much shorter than the finite radius of the
radiator are shown, see Eqs.(\ref{setteter},\ref{settequater}). In
Fig.(\ref{Fig1}), for a beam energy of $500 MeV$ and a bunch of
$N=10^5$ electrons, the angular distribution of the transition
radiation intensity in the visible (Green, $\lambda=530 nm$) for
different values of the beam standard deviation $\sigma=10, 50, 100$
$\mu m$ is calculated and compared with the ideal result obtainable
from a bunch with the same charge but radiating according to the
formula given in Eq.(\ref{nove}). In
Figs.(\ref{Fig2},\ref{Fig3},\ref{Fig4}), for different beam energies
$500, 750, 1000$ $MeV$ and a constant beam charge of $N=10^5$
electrons, the angular distributions of the transition radiation
intensity radiated at different wavelengths in the visible (Red,
Green, Blue $\lambda=680, 530, 400$ $nm$) by a beam with standard
deviation $\sigma=50$ $\mu m$ are compared with the analogous result
obtained in the ideal case from the formula in Eq.(\ref{nove}).

\section{Conclusions}
Temporal causality and covariance constrain the role that the
distributions of the longitudinal and transverse coordinates of an
electron bunch plays in determining the physical features of an
electromagnetic radiative mechanism, in particular, the transition
radiation mechanism. In the case of a $N$ electron bunch hitting at
a normal angle of incidence a metallic screen, the distribution of
the electron longitudinal coordinates defines the causal correlation
occurring between the temporal sequence of the electron collisions
onto the metallic screen and the structure of the relative emission
phases of the $N$ single electron radiation field amplitudes
propagating from the metallic screen to the observation point,
located on the beam longitudinal axis. The distribution of the
transverse coordinates of the $N$ electrons contributes as well to
determine the relative phase delay of the $N$ single electron
radiation field amplitudes at the observation point as a function of
the $N$ electron displacements with respect to the beam axis. The
transverse density of the $N$ electrons, being an invariant under a
Lorentz transformation with respect to the direction of motion of
the beam, constitutes for the radiation field a covariant feature
that affects both the spectral intensity and the angular
distributions of the transition radiation under observation
conditions of both temporal incoherence and temporal coherence.
Compared to the reference asymptotic results provided by the well
known ideal formalism of a single electron hitting an infinite
metallic screen, an increase of the radiated spectral intensity with
the decrease of the beam transverse size and a broadening of the
spectral angular distribution constitute the relevant
phenomenological aspects of the covariant beam-transverse-size
effects on the transition radiation energy spectrum originated by an
electron beam hitting a round metallic screen at a normal angle of
incidence.






\end{document}